\documentstyle[12pt,epsf]{article}   
\def\i{\item}
\def\ff{form factor }
\def\ffs{form factors }
\def\wt{Ward-Takahashi identity }
\def\cpt{$\chi$PT }
\def\beq{\begin{equation}}
\def\eeq{\end{equation}}
\def\ber{\begin{eqnarray}}
\def\eer{\end{eqnarray}}
\def\nn{\nonumber \\}
\def\ben{\begin{enumerate}}
\def\een{\end{enumerate}}
\def\bei{\begin{itemize}}
\def\eei{\end{itemize}}
\def\l{\label}
\def\s#1{#1\!\!/}
\def\se{self-energy }

\def\bi{\bibitem}
\def\nn{\nonumber \\}
\def\Lm{$\Lambda$ }

\def\Ia{\int d^4xe^{-i\ell\cdot x}\int d^4ye^{ip\cdot y}}
\def\Ib{\int d^4xe^{i\ell\cdot x}\int d^4ye^{-ip\cdot y}}
\def\dpl{\int \frac{d^4p}{(2\pi)^4}\int \frac{d^4\ell}{(2\pi)^4}}
\begin{document}
 
\begin{titlepage}
\begin{flushright} 
{nucl-th/9607022}
\end{flushright}
\vspace{0.5in}
\begin{Large}
\begin{center}
{\bf Meson-Baryon-Baryon Vertex Function and the Ward-Takahashi Identity}
\end{center}
\end{Large}
\vspace{5 mm}
\begin{large}
\begin{center}
{ Siwen Wang and Manoj K. Banerjee }
\end{center}
\end{large}
\vspace{5 mm}
\begin{center}
{ Department of Physics, University of Maryland, College Park, MD 20742}
\end{center}
\vspace{1.2in}

\begin{abstract}
Ohta proposed a solution for the well-known difficulty of satisfying the
Ward-Takahashi identity for a photo-meson-baryon-baryon amplitude
($\gamma$MBB)  when a dressed meson-baryon-baryon (MBB)  vertex
function is present.   He obtained a form for the  $\gamma$MBB  
amplitude which contained,  in addition to   the usual pole
terms, longitudinal seagull terms  which  were determined  entirely by the MBB
vertex function. He arrived at his result by using a Lagrangian which
yields the MBB vertex function at tree level.  We show that  such a
Lagrangian can be neither hermitian nor charge conjugation invariant. We
have been able to reproduce Ohta's result for the 
$\gamma$MBB  amplitude  using the Ward-Takahashi
identity and no other assumption, dynamical or otherwise,  and the most
general form for the MBB and $\gamma$MBB  vertices.  
However,  contrary to Ohta's finding, we  find that the   seagull
terms are not robust.
The seagull terms extracted from the  $\gamma$MBB  vertex  occur
unchanged in tree graphs, such as in an exchange current amplitude. But
the seagull terms which appear in a loop graph, as in the calculation
of  an  electromagnetic form factor, are, in general, different. 
The whole procedure says nothing about the
transverse part of the ($\gamma$MBB)  vertex and its contributions
to the amplitudes in question.
\end{abstract}
\vspace{0.5in}
\begin{center}
\end{center}
\end{titlepage}

\section{Introduction}
In a hadronic Lagrangian based approach to baryon properties
meson-baryon loops  must be calculated  to obtain the contribution of
virtual mesons.  In a wide variety of situations these loops diverge. 
There are two ways of dealing with the problem of divergence. One is
the well-established chiral perturbation theory ($\chi$PT)~\cite{CPT}, 
where only the pseudoscalar octet mesons are considered. The strategy 
is to use dimensional regularization and remove 
any  divergence arising from a loop with an appropriate
counterterm. The strength of the finite remainder, called a 
{\em low energy constant},  have to  be fixed with the help of
experimental data only. It may happen that one needs the physical
quantity under study itself to fix the low energy constant. In such a 
situations \cpt is unable to  make a prediction.

The other  popular approach is to introduce meson baryon \ffs to
regulate loop integrals. Unfortunately we know very little about such 
\ffs. The practice is to  extrapolate the limited information about the
\ff for a range of space-like meson  momentum, with the nucleon legs on 
mass-shell, to the full range of meson momentum using a multipole form. 
This very simple version of \ff  may be parameterized by one or two mass 
parameters. The masses are  guessed and are often assigned  values in 
the range $500$ MeV to $1400$MeV. Unfortunately one has difficulty 
dealing with the electromagnetic properties of the baryon when one uses 
a phenomenological \ff in general and  
approximates it with an algebraically convenient form in particular. The
electromagnetic vertex function $\Gamma_{\mu}^{em}(p+q,p)$, calculated
at the one-loop level, as shown in Fig.~\ref{V1}, 
\begin{figure}[h]
\hskip 0.75in
\epsfxsize=4.8in
\epsfysize=0.9in
\epsffile{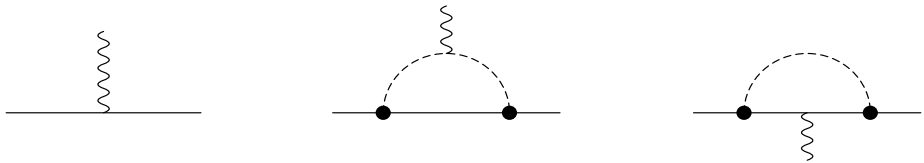}
\caption{The bare and one-loop contributions to the baryon
electromagnetic vertex function.  The MBB vertices are dressed.}
\l {V1}
\end{figure}
but using \ffs at the meson-nucleon vertices will not satisfy the
Ward-Takahashi identity:
\beq (p^\mu+q^\mu-p^\mu)\Gamma_{\mu}^{em}(p+q,p)=\s p+\s q-\s p-
\Sigma(p+q)+\Sigma(p), 
\l {wt1} \eeq 
where $\Sigma(p)$ is the proper self-energy, $p$ and $p+q$ are the initial 
and final momenta. In fact,   one does not even know what the expression
for the self-energy is when a 
parameterized  \ff is used without any dynamical basis.  

Ohta ~\cite{KO} proposes a particular path to resolving this problem. 
There are four features in his approach.
\bei
\i He introduces an interaction Lagrangian which reproduces the
meson-baryon vertex function at the tree level.  Naturally the
Lagrangian uses functions of the derivative operators which operates on
the fields associated with the three external legs of the vertex. 
\i The electromagnetic interactions are generated via minimal coupling:
$\partial_\mu\rightarrow \partial_\mu+iQA_\mu$, where $Q$ is the  
operator which measures the charge of the field on which $\partial_\mu$ 
operates.
\i He then notes that the photo-meson amplitude resulting from his
procedure contained a longitudinal seagull term\footnote{ A seagull
term cannot be cut into a MBB vertex  and a  $\gamma$MM 
($\gamma$BB vertex) by cutting a single meson   (baryon) propagator.} 
expressible entirely in terms of the vertex function. The presence of this  
longitudinal seagull term is essential for satisfying the \wt. 
\i  The  seagull term found by Ohta is  robust in the context of his
theory.  The same seagull term appears in any meson-baryon one loop
graph where the photon interacts with an internal line . Also the  
seagull term appears at each meson-baryon  vertex in the loop, thus
generating two graphs.
\eei

Because of its simplicity Ohta's prescription has become quite popular.
Several authors have used it for one loop calculations of baryon
electromagnetic  and strangeness
properties~\cite{musolf,musolf2,cohen,forkel,koepf,frank}.

We  show that Ohta's Lagrangian is necessarily neither hermitian
nor invariant under charge conjugation. Yet we succeed in reproducing Ohta's
prescription  by using only the Lorentz structure of the $\gamma$MBB
vertex and the Ward-Takahashi identity itself.

Unfortunately, we also find that,  in general, the resulting seagull
term is not robust.   It should be noted that in a tree graph with
dressed vertices as in the exchange current graphs, the seagull terms are
exactly those appearing  in a photo-meson amplitude. A loop graph of an
electromagnetic amplitude will contain a seagull term if a meson baryon form 
factor has been used, but, in general, it will not be the same as that 
appearing in a photo-meson amplitude.

The absence of robustness is established  with the help of specially chosen
subsets of Feynman graphs for   KP$\Lambda$  and  $\gamma$KP$\Lambda$
vertices.  After the seagull terms in the $\gamma$KP$\Lambda$ vertex
are identified, we check their robustness by considering the one loop
graphs for the strangeness content of proton.  As we have noted, the
seagull terms act robustly in dressed tree graphs.  We consider two
examples. In one, the lowest order (Born) graphs are dressed only by
kaon self-energy. This does generate robust seagull terms.  In the
other the dressing is provided by ladders of a neutral scalar meson 
which couples only to strangeness. The resulting seagull terms are not
robust. In fact the one loop graphs for the strangeness content of
proton in this case does not even have any  seagull term. This example
is sufficient to establish that, in general,  the seagull terms
identified from  the $\gamma$MBB vertex are not robust.
 
The paper is organized as follows. The next section describes briefly
Ohta's strategy and mentions the difficulties.  Section~\ref{hermit} 
discusses  the lack of hermiticity and   the absence of charge
conjugation invariance of the Ohta Lagrangian.  In  section 4 we
analyze   general 3-point and 4-point functions related to
MBB   and   $\gamma$MBB vertices and identify the seagull terms
occurring in the latter.  The result  matches Ohta's prescription. The
only underlying assumption is that Lorentz  and translational
invariances are good symmetries of the theory. Otherwise, there is no
reference to any particular dynamical model. 

The robustness of the resulting seagull terms is discussed in
section~\ref{robust}.  Section~\ref{rescon} contains a summary of the 
results and conclusions.

\section{The Ohta strategy and difficulty} \label{ohta}
We mentioned in the previous section that the  choice of  a convenient
form for the \ff tells us nothing about the self-energy, thus making it 
impossible 
even to discuss the question of the Ward-Takahashi identity. One needs
to know  the Lagrangian
which generates the \ff and the self-energy.  Ohta~\cite{KO} solved the  
problem by writing down an interaction Lagrangian which reproduces the 
arbitrarily guessed \ff at the {\em tree} level. Here we summarize his work 
very briefly. To appreciate his work fully one must read Ref.~\cite{KO}.  

We also simplify the presentation by substituting the more realistic
pseudoscalar meson field used by Ohta with a scalar field. While the
substitution halves the number of independent covariant forms needed to
describe the vertex function, it does not affect the thrust of this paper.

In  Ohta's approach  the scalar  meson-baryon-baryon \ff , $\Gamma(\ell,p)$ 
is reproduced at tree level by the interaction Lagrangian \beq
{\cal L}(x',x,y)=\bar{\psi}(x')\Gamma(x',x,y)\psi(x)\phi(y). \l {OL1} \eeq
It should be noted that in this approach there are no self-energy
insertions on the meson and baryon lines.  Hence there are no factors
contributing  to the vertex functions  from wave function
renormalizations. The quantity $\Gamma(x',x,y)$ is the full vertex function.
The Fourier transform of  the nonlocal function $\Gamma(x',x,y)$ gives 
the MBB vertex function $\Gamma(\ell,p)$ as follows 
\beq
\Gamma(x',x,y)=\int\frac{d^4 p}{(2\pi)^4}\frac{d^4 \ell}{(2\pi)^4}
e^{i\ell \cdot(x'-y)-ip\cdot(y-x)}\Gamma(\ell,p), 
\l {OL2} \eeq
where $p$ and $\ell$ are the initial and final baryon momenta. One can expand 
the vertex function $\Gamma(\ell,p)$ in terms of a complete set of
appropriate Lorentz covariant forms.   A perfectly  general form for a
scalar meson MBB vertex is
\begin{eqnarray}
\Gamma(\ell,p)&=&g_0(k^{2},\ell^{2},p^{2})+g_1(k^{2},\ell^{2},p^{2})(\s
\ell -\s p)+g_2(k^{2},\ell^{2},p^{2})(\s
\ell +\s p) \nonumber \\
&+& g_3(k^{2},\ell^{2},p^{2})[\s \ell ,\s p], \label{Gamex} \\
k&=&p-\ell. \label{Gamex1} 
\end{eqnarray}
where $ g_i(k^{2},\ell^{2},p^{2})$  are  Lorentz scalar functions. 
Following Ohta, we  introduce a 
generalized function  of three independent momenta $\bar{\Gamma}(k,\ell,p)$
by removing the restriction imposed by Eq.~(\ref{Gamex1}).
\begin{eqnarray}
\bar{\Gamma}(k,\ell,p)&=&g_0(k^{2},\ell^{2},p^{2})+g_1(k^{2},\ell^{2},p^
{2})(\s \ell -\s p)+g_2(k^{2},\ell^{2},p^{2})(\s
\ell +\s p) \nonumber \\
&+& g_3(k^{2},\ell^{2},p^{2})[\s \ell ,\s p], \label{Gmab} \\
k&\neq&p-\ell, {\rm\,\,in\,\,general,} \label{Gmab1}  
\end{eqnarray}
and 
\begin{equation}
\bar{\Gamma}(k=p-\ell,\ell,p)=\Gamma(\ell,p),~~~~{\rm in ~particular}. 
\label{Gmab2}
\end{equation}
The definition of $\bar{\Gamma}(k,\ell,p)$ requires the knowledge of the
analytic structure of the scalar functions
$g_i=g_i(k^{2},\ell^{2},p^{2})$ which appear in Eq.~(\ref{Gamex}).
Note also that if we had chosen $(\ell, k)$ or $(k,p)$ as the
independent pair of momenta in Eq.~(\ref{Gamex}), the resulting
analytically continued object $\bar{\Gamma}(k,\ell,p)$ would be
different. However, Eq.~(\ref{Gmab2}) will continue to hold.

Ohta then introduces the electromagnetic interaction via minimal
substitution. The procedure yields  new seagull terms whose structure
depends upon the quantities  $\bar{\Gamma}(k,\ell,p)$. The explicit form of 
the seagull term
\begin{figure}[h]
\hskip 2.3in
\epsfxsize=1.6in
\epsfysize=0.8in
\epsffile{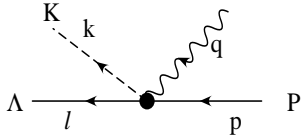}
\caption{ The KP\Lm vertex and Ohta's seagull term for the photo-kaon
production amplitude.}
\l {fig_ohta1} 
\end{figure} 
for photo-kaon production amplitude, shown in Fig.~\ref{fig_ohta1}, is 
given below.
\ber
{\cal M}^{sg}_\mu &=&-Q_K\frac{(2k-q)_\mu}{k^2-(k-q)^2}
[\bar{\Gamma}(k,\ell,p)-\bar{\Gamma}(k-q,\ell,p)] \nn
&& -Q_\Lambda\frac{(2\ell-q)_\mu}{\ell^2-(\ell-q)^2}
[\bar{\Gamma}(k,\ell,p)- \bar{\Gamma}(k,\ell-q,p)] \nn
&&
-Q_P\frac{(p+q+p)_\mu}{(p+q)^2-p^2}[\bar{\Gamma}(k,\ell,p+q)-
\bar{\Gamma}(k,\ell,p)] 
+{\cal M}^{t}_\mu,
\l {OL5} 
\eer
where ${\cal M}^{t}_\mu$ is some   transverse component of ${\cal
M}^{sg}_\mu$.  Note that  $k\neq p-\ell$ in
the term  $\bar{\Gamma}(k,\ell, p)$ which appears in Eq.~(\ref {OL5}). 
Note, further, that the sum of the three terms containing this factor
is diveregenceless and is thus   a transverse term.\footnote{ 
Ohta gave the  explicit form of ${\cal M}^{t}_\mu$ for the 
pseudoscalar  meson production resulting from  his dynamical model.} 

Since these terms arise from a standard method of adding electromagnetic 
interaction to a strong interaction Lagrangian, unusual as the latter may be, 
it is not surprising these seagull terms, together with the usual generalized 
Born terms containing either a kaon pole or a baryon pole, satisfy the
Ward-Takahashi identity.
 
Unfortunately there is a serious  problem with Ohta's prescription. The
Lagrangian~(\ref{OL1}) is not hermitian   if the MBB vertex 
$\Gamma(\ell,p)$ originally came from some Lagrangian satisfying the 
standard
requirements of invariance under Lorentz transformation, time reversal,
parity and charge conjugation. The basis of our claim is described in
the next section.

\section{Hermiticity and charge conjugation invariance of the Ohta
Lagrangian} \label{hermit}
The baryon and kaon propagators are defined as follows. The Feynman
propagators with renormalized masses are:
\begin{eqnarray} 
S^0(p)&=&\frac{1}{\s p-M}
\nonumber \\
\Delta^0(k)&=& \frac{1}{k^2-m^2},
\label{eq:baressk}
\end{eqnarray}
and the dressed propagators are 
\begin{eqnarray}
S(p)&=&\frac{1}{\s p-M-\Sigma(p)}=\tilde{F}(p)S^0(p)\tilde{F}(p)
\nonumber \\
\Delta(k)&=& \frac{1}{k^2-m^2-\Sigma(k)}
=\tilde{f}(k^2)\Delta^0(k)\tilde{f}(k^2),
\label{eq:ssk}
\end{eqnarray}
where the functions $\tilde{F}(p)= A(p^2)+\s p B(p^2)$ and $\tilde{f}(k^2)$ 
are introduced to incorporate the contribution of wave function
renormalizations  to the general vertex function. $A(p^2)$, $B(p^2)$
and $\tilde{f}(k^2)$ can be expressed in terms of the baryon and meson
self-energies.

\begin{figure}[h]
\hskip 1.25in
\epsfxsize=4.5in
\epsfysize=0.9in
\epsffile{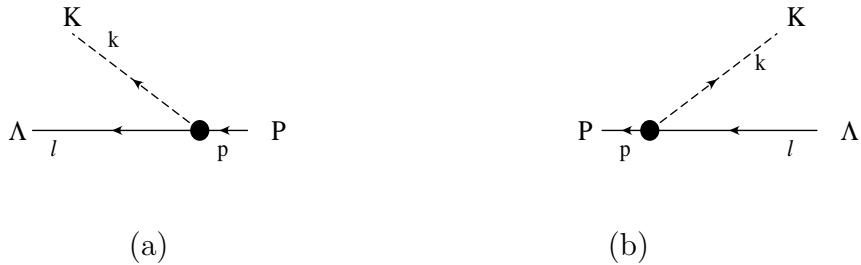}
\vskip 0.25in
\centerline{(a)\hspace{2.3in}(b)}
\caption{ (a) corresponds to proton becoming $\Lambda$ and $K^+$.
The associated vertex function is $\Gamma(\ell,p)$.    (b)
correspond to $\Lambda$ becoming proton and $K^-$. The associated
vertex function is $ \Gamma'(p,\ell)$.}
\label {fig1}
\end{figure}

Next we introduce the
three-point functions, $G(\ell,p)$ and $G'(p,\ell)$ and the vertex
functions, $\Gamma(\ell,p)$ and
$\Gamma'(p,\ell)$ corresponding to the graphs as shown in Fig. \ref{fig1}.
\begin{eqnarray}
G(\ell,p)&=&\Ib\langle 0\mid
T(\psi^{\Lambda}(x)\phi(0)\bar{\psi}^P(y))\mid 0\rangle \nn
&=&
-i\tilde{F}_\Lambda(\ell)S^0_{\Lambda}(\ell)\Gamma(\ell,p)S^0_P(p)\tilde
{F}_P(p)\Delta^0(p-\ell)\tilde{f}((p-\ell)^2),\label {gam} \\
G'(p,\ell)&=&\Ia\langle 0\mid
T(\psi^P(y)\phi^*(0)\bar{\psi^\Lambda}(x))\mid 0\rangle \nn
&=&
-i\tilde{F}_P(p)S^0_P(p)\Gamma'(p,\ell)S^0_{\Lambda}(\ell)
\tilde{F}_{\Lambda}(\ell)\Delta^0(\ell-p)\tilde{f}((p-\ell)^2).
\label {gamp} 
\end{eqnarray}
We take it that the vertex functions have been generated by a Lagrangian 
which is invariant under charge conjugation. Naturally, the resulting
vacuum is even under charge conjugation. Exploiting this fact we find that
\begin{eqnarray}
G_{\alpha,\beta}(\ell,p)&=& \Ib\langle{\cal C} 0\mid 
T(\psi^\Lambda_\alpha(x)\phi(0)\bar{\psi}^P_\beta(y))
\mid{\cal C} 0\rangle \nonumber \\ 
&=& \Ib\langle 0\mid {\cal C}
T(\psi^\Lambda_\alpha(x)\phi(0)\bar{\psi}^P_\beta(y))
{\cal C}^{-1}\mid 0\rangle \nonumber \\ 
&=& \Ib C_{\alpha\alpha'}\langle 0\mid T(\psi^P_{\beta'}(y) \phi^*(0)
\bar{\psi}^\Lambda_{\alpha'}(x))\mid 0\rangle C^{-1}_{\beta'\beta},
\end{eqnarray}
where the matrix $C$ is given by 
\begin{equation}
C=i\gamma^2\gamma^0=-C^{-1}. 
\label{C1} 
\end{equation} 
By using the definition of $G'(p,\ell)$, given by Eq. (\ref{gamp}) we obtain  
\begin{eqnarray}
G_{\alpha,\beta}(\ell,p)&=& \Ib C_{\alpha\alpha'}\langle 0\mid 
T(\psi^P_{\beta'}(y)\phi^*(0)\bar{\psi}^\Lambda_{\alpha'}(x))
\mid 0\rangle C^{-1}_{\beta'\beta} \nonumber \\ 
&=&C_{\alpha\alpha'}G'_{\beta',\alpha'}(-p,-\ell)
C^{-1}_{\beta'\beta}\nonumber \\ 
&=&(C G'^{\rm T}(-p,-\ell)C^{-1})_{\alpha\beta},\\
\label{cgcx}
\end{eqnarray}
that is,
\begin{equation}
G(\ell,p)=CG'^{\rm T}(-p,-\ell)C^{-1}, \label {cgc} \end{equation}
where the superscript T is for the transpose.   
The fact that $C\s p^T C^{-1}=-\s p$ leads to results
\begin{equation}
 C S^0(-p)^T C^{-1}=S^0(p)\,\,\,\,{\rm and}\,\,\,\,
C\tilde{F}(-p)^T C^{-1}=\tilde{F}(p). 
\label {SC}  
\end{equation}   
Using the above equation and Eqs.(\ref{cgc}) and  (\ref{gamp}) we   have
\begin{equation}
\Gamma(\ell,p)=C\Gamma'^{\rm T}(-p,-\ell)C^{-1}. 
\label {GC}
\end{equation}
 
We recall the expansion of $\Gamma(\ell,p)$, given by Eq.~(\ref{Gamex})
and supplement it with the corresponding expansion of $\Gamma'(p,\ell)$:
\begin{eqnarray}
\Gamma(\ell,p)&=&g_0(k^{2},\ell^{2},p^{2})+g_1(k^{2},\ell^{2},p^{2})(\s
\ell -\s p)+g_2(k^{2},\ell^{2},p^{2})(\s \ell +\s p)\nonumber \\
&+& g_3(k^{2},\ell^{2},p^{2})[\s \ell ,\s p], \nonumber \\
 \Gamma'(p,\ell)&=&
g'_0(k^{2},p^{2},\ell^{2})+g'_1(k^{2},p^{2},\ell^{2})(\s p -\s
\ell)+g'_2(k^{2},p^{2},\ell^{2})(\s p+\s \ell)\nonumber \\
&+&g'_3(k^{2},p^{2},\ell^{2})[\s p,\s \ell]. 
\label{texGampex} 
\end{eqnarray}

Combining Eq.~(\ref{GC}) with these expansions we    find the following 
symmetry relations between the two sets of Lorentz scalar functions $g_i$ 
and $g'_i$
\begin{eqnarray}
g_0(k^{2},\ell^{2},p^{2})&=& g'_0(k^{2},p^{2},\ell^{2}), \nonumber \\
g_1(k^{2},\ell^{2},p^{2})&=&-g'_1(k^{2},p^{2},\ell^{2}), \nonumber \\
g_2(k^{2},\ell^{2},p^{2})&=& g'_2(k^{2},p^{2},\ell^{2}), \nonumber \\
g_3(k^{2},\ell^{2},p^{2})&=& g'_3(k^{2},p^{2},\ell^{2}), 
\label{gigpi} 
\end{eqnarray}

The  Ohta  
Lagrangian which  generates the two
vertex functions of Fig.~(\ref{fig1}) at the tree level is the following:
\begin{eqnarray}
{\cal L}_{int}&=&\dpl e^{i\ell\cdot x-ip\cdot y +i(p-\ell)\cdot z} 
\bar{\psi}^{\Lambda}(x)\Gamma(\ell,p)\phi(z)\psi^{P}(y) \nonumber \\
&+&\dpl e^{-i\ell\cdot x +ip\cdot y-i(p-\ell)\cdot z}
\bar{\psi}^{P}(y) \Gamma'(p,\ell)\phi^*(z)\psi^{\Lambda}(x).
\label{texLag}
\end{eqnarray}
 
The requirement that ${\cal L}_{int}$ be hermitian  demands that
\begin{equation}
\Gamma'(p,\ell)=\gamma_0\Gamma^\dagger(\ell,p)\gamma_0. \label
{texGherm}\end{equation}
Using the above equation in conjunction with the expansions given in
Eq.~(\ref{texGampex}) and the relations of Eq.~(\ref{gigpi})
 we find that the
hermiticity of the Ohta's Lagrangian demands that 
\begin{equation}
g_{i}(k^{2},\ell^{2},p^{2})= g_{i}^{*}(k^{2},\ell^{2},p^{2}).
\label{gprop}
\end{equation}
In other words, hermiticity of the interaction Lagrangian combined with
charge-conjugation invariance of the fundamental dynamics demands that
the coefficient functions be always real. However, it is well known that 
they are necessarily complex when $\ell^2$, $p^2$ and
$(p-\ell)^2$ have appropriate values. Thus for example, when
$\ell^2=M_\Lambda^2$, $p^2=M_P^2$ and $(p-\ell)^2\geq 9m^2$ the
functions $g_{i}$ will become complex. In a realistic situation with pions
present the analytic structure is richer with threshold occurring at
lower masses. The reader may also verify the essential points of the
preceding remarks by studying the vertex function at one-loop level of 
any form.

\section{The  \wt and $\gamma$MBB vertex}\label{wt}
In this section we reproduce Ohta's final result for $\gamma$MBB 
vertex using solely the \wt and the freedom to add
divergenceless, {\it i.e.,} transverse, parts to the expressions. No
reference is made to any particular Lagrangian or dynamical model. 

At the center is the four-point function $\langle T(\psi^\Lambda(x')
\bar{\psi}^P(x)\phi(y)J_{\mu}(z))\rangle$, where $J_{\mu}(z)$ is the
electromagnetic current.   The fields satisfy the equal time commutation 
relations:
\begin{eqnarray}
\delta(x_0-y_0)[J_0(x),\psi(y)]&=&-\delta^4(x-y)Q_{\psi}\psi(x), \nonumber \\
\delta(x_0-y_0)[J_0(x),\phi(y)]&=&-\delta^4(x-y)Q_{\phi}\phi(x),
\label{charge} \end{eqnarray} 
where $Q_{\psi}$ and $Q_{\phi}$ are the charges of the particles
described by the fields.
Using these  equal time commutation relations one obtains the following
expression for the divergence of the four-point function:
\begin{eqnarray}
\partial^{\mu}_{z}\langle T(\psi^{\Lambda}(x')\bar{\psi}^P(x)
\phi(y)J_{\mu}(z))\rangle & = & -[Q_{\Lambda}\delta^{4}(x'-z)
-Q_{P}\delta^{4}(x-z)+Q_{K}\delta^{4}(y-z)] \nonumber \\
&  & \times \langle T(\psi^\Lambda(x')
\bar{\psi}^P(x)\phi(y))\rangle
\label{eq:wt1p}
\end{eqnarray} 

The $\gamma$MBB vertex function, represented by the first graph in
Fig.~\ref{fig_4pt}  is defined by the equation
\begin{eqnarray}
&&i\int d^4xe^{i\ell\cdot x}\int d^4x'e^{-ip\cdot x'}\int
d^4ye^{ik\cdot y}\langle T(\psi^\Lambda(x')\bar{\psi}^P(x)
\phi(y)J_{\mu}(0))\rangle \nonumber \\
&=&  \tilde{F}_{\Lambda}(\ell)S^0_{\Lambda}(\ell)\tilde{f}(k^2)\Delta^0(k)
{\cal M}_{\mu}(k,\ell,p)S^0_{P}(p)\tilde{F}_{P}(p),
\label{newm2} 
\end{eqnarray}
Using Eqs. (\ref{gam}), (\ref{eq:wt1p}) and (\ref{newm2}) and after 
amputating 
the external factor $\tilde{F}_{\Lambda}(\ell)S^0_{\Lambda}(\ell)$, 
$S^0_{P}(p)\tilde{F}_{P}(p)$ and $\Delta^0(k)\tilde{f}(k^2)$, we obtain 
the Ward-Takahashi identity: 
\begin{eqnarray}
q^{\mu}\cdot {\cal M}_{\mu} &=& Q_{K}\Delta^{0}(k)^{-1}\tilde{f}^{-1}
(k^2)\Delta^{0}(k-q)\tilde{f}((k-q)^2)\Gamma(\ell,p)\nonumber \\
&-& Q_{P}\Gamma(\ell,p+q)S^0_P(p+q)\tilde{F}_{P}(p+q)
\tilde{F}^{-1}_{P}(p)S^0_{P}(p)^{-1} \nonumber \\
&+& Q_{\Lambda}S^0_{\Lambda}(\ell)^{-1}\tilde{F}^{-1}_{\Lambda}(\ell)
\tilde{F}_{\Lambda}(\ell-q)S^0_{\Lambda}(\ell-q)\Gamma(\ell-q,p).
\label{eq:wt2} 
\end{eqnarray}

Following the standard practice, we  express ${\cal M}_\mu$ as a
sum of three `pole' terms, defined by us, and the remainder, which is
free of meson or baryon poles. This is exhibited in Fig.~\ref{fig_4pt}.
Thus by definition the last term  is the seagull term.  A further
refinement of the definition of the `seagull' term will follow. The
`pole' terms are not pure poles as they contain the full MBB vertex
function.  A pure pole term will have   the appropriate momentum  in
the vertex function on its mass shell.  The electromagnetic vertices in
the pole terms are just the lowest order terms with renormalized
charges. Thus the photo-kaon vertex is
$-iQ_K\Gamma^{(K)em}_\mu(k,k-q)=-iQ_K(2k-q)_\mu$.  
\begin{figure}[h]
\hskip 0.25in
\epsfxsize=6in
\epsfysize=0.78in
\epsffile{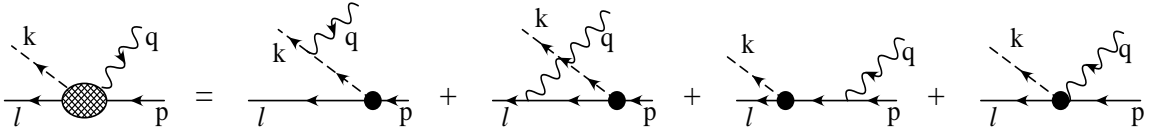}
\caption{The $\gamma$MBB vertex ${\cal M}_{\mu}(k,\ell,p) =$ 
three pole terms + a seagull part }
\l {fig_4pt} 
\end{figure}
The expression for  ${\cal M}^{pole}_{\mu}$,  the sum of
the three pole terms of Fig.~\ref{fig_4pt} with the renormalized charges, is 
\begin{eqnarray}
{\cal M}^{pole}_{\mu} &=& Q_{K}
\Gamma_{\mu}^{(K)em}(k,k-q)\Delta^0(k-q)\Gamma(\ell,p) \nonumber \\
&+& Q_{\Lambda}\Gamma_{\mu}^{(\Lambda)em}(\ell,\ell-q)
S^0_{\Lambda}(\ell-q)\Gamma(\ell-q,p) \nonumber \\
&+&Q_P \Gamma(\ell,p+q)S^0_{P}(p+q)\Gamma_{\mu}^{(P)em}(p+q,p),
\label{eq:poledef}
\end{eqnarray}
The Ward-Takahashi identities for these vertices are, 
\begin{eqnarray}
q^{\mu}\cdot \Gamma_{\mu}^{(K)em}(k,k-q) &=&
\Delta^0(k)^{-1}-\Delta^0(k-q)^{-1} \nonumber \\
q^{\mu}\cdot \Gamma_{\mu}^{(\Lambda)em}(\ell,\ell-q) &=& 
S^0_{\Lambda}(\ell)^{-1}-S^0_{\Lambda}(\ell-q)^{-1} \nonumber \\
q^{\mu}\cdot \Gamma_{\mu}^{(P)em}(p+q,p) &=& 
S^0_{P}(p+q)^{-1}-S^0_{P}(p)^{-1} \label{eq:wtcurr}
\end{eqnarray}
The divergence of the ${\cal M}^{pole}_{\mu}$ is
\begin{eqnarray}
q^{\mu}\cdot {\cal M}^{pole}_{\mu} &=& 
Q_{K}[\Delta^0(k)^{-1}\Delta^0(k-q)-1]\Gamma(l,p)\nonumber \\
&+&Q_{P}\Gamma(\ell,p+q)[1-S^0_{P}(p+q)S^0_{P}(p)^{-1}] \nonumber \\
&+& Q_{\Lambda}[S^0_{\Lambda}(\ell)^{-1}S^0_{\Lambda}(\ell-q)-1]
\Gamma(\ell-q,p)
\label{eq:pole}
\end{eqnarray}
By definition
\begin{equation}
{\cal M}^{sg}_\mu={\cal M}_{\mu}- {\cal M}^{pole}_{\mu}
\end{equation}
 From Eqs. (\ref{eq:wt2}) and (\ref{eq:pole}) we have
\begin{eqnarray}
q^{\mu}\cdot {\cal M}^{sg}_{\mu} &=& 
q^{\mu}\cdot({\cal M}_{\mu}- {\cal M}^{pole}_{\mu}) \nonumber \\
&=& Q_{K}\Gamma(\ell,p)[1-\Delta^0(k)^{-1}\tilde{f}^{-1}(k^2)
\Delta^0(k-q)(\tilde{f}(k^2)-\tilde{f}((k-q)^2))] \nonumber \\
&+& Q_{\Lambda}[1-S^0_{\Lambda}(\ell)^{-1}\tilde{F}^{-1}_{\Lambda}
(\ell)(\tilde{F}_{\Lambda}(\ell)-\tilde{F}_{\Lambda}(\ell-q))
S^0_{\Lambda}(\ell-q)]\Gamma(\ell-q,p) \nonumber \\
&-& Q_{P}\Gamma(\ell,p+q)[1-S^0_{P}(p+q)(\tilde{F}_{P}(p)-
\tilde{F}_{P}(p+q))\tilde{F}^{-1}_{P}(p)S^0_{P}(p)^{-1}]. 
\label{eq:seagul}
\end{eqnarray}
The vertex functions in Eq. (\ref{eq:seagul}) all satisfy 
momentum conservation.  For future convenience in matching our result with 
that of Ohta we write these as 
$\bar{\Gamma}(k-q,\ell,p)$,  etc.  using  
Eq. (\ref{Gmab2}). The longitudinal part of ${\cal M}^{sg}_{\mu}$ is fixed by 
the \wt and we may write the solution of Eq.~(\ref{eq:seagul}) in the form:  
\begin{eqnarray}
{\cal M}^{sg}_\mu &=&Q_K\frac{(2k-q)_\mu}{k^2-(k-q)^2}
\bar{\Gamma}(k-q,\ell,p)[1-\Delta^0(k)^{-1}\tilde{f}^{-1}(k^2)\Delta^0(k-q)
(\tilde{f}(k^2)-\tilde{f}((k-q)^2))] \nonumber \\
&+&Q_\Lambda\frac{(2\ell-q)_\mu}{\ell^2-(\ell-q)^2}
[1-S^0_{\Lambda}(\ell)^{-1}\tilde{F}^{-1}_{\Lambda}(\ell)
(\tilde{F}_{\Lambda}(\ell)-\tilde{F}_{\Lambda}(\ell-q))
S^0_{\Lambda}(\ell-q)]\Gamma(\ell-q,p) \nonumber \\
&-&Q_P\frac{(p+q+p)_\mu}{(p+q)^2-p^2}\bar{\Gamma}(k,\ell,p+q)
[1-S^0_{P}(p+q)(\tilde{F}_{P}(p)-\tilde{F}_{P}(p+q))
\tilde{F}^{-1}_{P}(p)S^0_{P}(p)^{-1}]
\nonumber \\
&+&{\cal M}^{t}_{\mu},
\label{Msg1}
\end{eqnarray}
where ${\cal M}^{t}_{\mu}$ is an unknown  transverse part. The
preceding expression contains many pieces involving propagators and
wave function renormalization factors which do not appear in Ohta's
treatment. The reason is that the latter is based on a very special
dynamics while the result here is completely general. However, we note
that these extra terms all vanish when the external legs are on their
respective mass shells, {\it i.e.,} $k^2=m^2$ and one   replaces   
$\s p $ with $M_P$ and 
$\s \ell$ with $M_{\Lambda}$. Exploiting this we propose that the
general result should be compared with Ohta's on the mass shell only.
Whereupon we have
 \begin{eqnarray}
{\cal M}^{sg}_\mu &=&Q_K\frac{(2k-q)_\mu}{k^2-(k-q)^2}
\bar{\Gamma}(k-q,\ell,p)+Q_\Lambda\frac{(2\ell-q)_\mu}{\ell^2-(\ell-q)^2}
\bar{\Gamma}(k,\ell-q,p) \nn
&& -Q_P\frac{(p+q+p)_\mu}{(p+q)^2-p^2}\bar{\Gamma}(k,\ell,p+q)
+{\cal M}^{t}_{\mu}.
\label{Msg}
\end{eqnarray}
The solution   still does not quite match Ohta's result.  It also has
the unpleasant feature that it is singular when either $q\rightarrow 0$
or $k^2 \rightarrow (k-q)^2$ etc. We rectify the problem by choosing 
following form for  ${\cal M}^{t}_{\mu}$:
\begin{eqnarray}
{\cal M}^{t}_{\mu} = {\cal M}^{t}_{1,\mu}+ {\cal M}^{t}_{2,\mu}+ 
{\cal M}^{t}_{3,\mu}, 
\label{mt12}
\end{eqnarray}
where
\begin{eqnarray}
{\cal M}^{t }_{1,\mu}&=&-Q_K\frac{(2k-q)_\mu}{k^2-(k-q)^2}
\bar{\Gamma}(k,\ell,p)
-Q_\Lambda\frac{(2\ell-q)_\mu}{\ell^2-(\ell-q)^2}
\bar{\Gamma}(k,\ell,p) \nn
&& +Q_P\frac{(p+q+p)_\mu}{(p+q)^2-p^2} \bar{\Gamma}(k,\ell,p), 
\label{mt1}
\end{eqnarray}  
which ensures the finiteness of ${\cal M}^{sg}_\mu$ as $q\rightarrow 0$
or $k^2=(k-q)^2$, and
\begin{eqnarray} 
{\cal M}^{t}_{2,\mu} &=& Q_{\Lambda}\{\frac{(2\ell-q)_{\mu}}
{\ell^2-(\ell-q)^2} q\!\!\!/-\gamma_{\mu}\}
[g_1(k^2,(\ell-q)^2,p^2)+g_2(k^2,(\ell-q)^2,p^2)] \nn
&& +Q_{\Lambda}\{\frac{(2\ell-q)_{\mu}}{\ell^2-(\ell-q)^2}[q\!\!\!/,p\!\!\!/]
+2i\sigma_{\mu \nu}p^{\nu}\}g_3(k^2,(\ell-q)^2,p^2) \nn
&& + Q_P\{\frac{(p+p+q)_{\mu}}{(p+q)^2-p^2}q\!\!\!/-\gamma_{\mu}\}
[g_1(k^2,\ell^2,(p+q)^2)+g_2(k^2,\ell^2,(p+q)^2)] \nn
&& + Q_P\{\frac{(p+p+q)_{\mu}}{(p+q)^2-p^2}[\ell\!\!\!/,q\!\!\!/]
+2i\sigma_{\nu \mu}\ell^{\nu}\}g_3(k^2,\ell^2,(p+q)^2),
\label{mt2}
\end{eqnarray}
which guarantees the regularities of ${\cal M}^{sg}_\mu$  when either
$\ell^2=(\ell-q)^2$ or $(p+q)^2=p^2$. The quantity 
${\cal M}^{t}_{3,\mu}$ is still some undetermined transverse term which is
however free from unphysical singularities.

Keep in mind that while 
$\bar{\Gamma}(k-q,\ell,p)=\Gamma(\ell,p)$, 
$\bar{\Gamma}(k,\ell-q,p)=\Gamma(\ell-q,p)$
and $\bar{\Gamma}(k,\ell,p+q)=\Gamma(\ell,p+q)$ are true vertex functions
satisfying the momentum conservation, $\bar{\Gamma}(k,\ell,p)$,
which is defined {\em via} analytical continuation of $g_i$ etc in 
Eqs. (\ref{Gmab}) and (\ref{Gmab1}), is equal to a  physical MBB 
vertex at $q=0$ only. 

Eqs. (\ref{Msg}), (\ref{mt1}) and (\ref{mt2}) together give us the
following result for the seagull vertex  
\begin{eqnarray}
{\cal M}^{sg}_\mu &=& -Q_K\frac{(2k-q)_\mu}{k^2-(k-q)^2}
[\bar{\Gamma}(k,\ell,p)-\bar{\Gamma}(k-q,\ell,p)] \nn
&& -Q_\Lambda\frac{(2\ell-q)_\mu}{\ell^2-(\ell-q)^2}
[\bar{\Gamma}(k,\ell,p)- \bar{\Gamma}(k,\ell-q,p)] \nn
&& -Q_P\frac{(p+q+p)_\mu}{(p+q)^2-p^2}
[\bar{\Gamma}(k,\ell,p+q)-\bar{\Gamma}(k,\ell,p)] \nn
&& +Q_{\Lambda}\{\frac{(2\ell-q)_{\mu}}{\ell^2-(\ell-q)^2} 
q\!\!\!/-\gamma_{\mu}\}[g_1(k^2,(\ell-q)^2,p^2)+
g_2(k^2,(\ell-q)^2,p^2)] \nn
&& +Q_{\Lambda}\{\frac{(2\ell-q)_{\mu}}{\ell^2-(\ell-q)^2}[q\!\!\!/,p\!\!\!/]
+2i\sigma_{\mu \nu}p^{\nu}\}g_3(k^2,(\ell-q)^2,p^2) \nn
&& + Q_P\{\frac{(p+p+q)_{\mu}}{(p+q)^2-p^2}q\!\!\!/-\gamma_{\mu}\}
[g_1(k^2,\ell^2,(p+q)^2)+g_2(k^2,\ell^2,(p+q)^2)] \nn
&& + Q_P\{\frac{(p+p+q)_{\mu}}{(p+q)^2-p^2}[\ell\!\!\!/,q\!\!\!/]
+2i\sigma_{\nu \mu}\ell^{\nu}\}g_3(k^2,\ell^2,(p+q)^2).
\nn
&& +{\cal M}^{t}_{3,\mu}.
\label{finmsg}
\end{eqnarray}

We find that
Eq.~(\ref{finmsg}) would be  exactly equal to Ohta's result had he discussed 
the case of a scalar meson instead of a pseudoscalar meson and chosen to 
expand the vertex function $\Gamma(\ell,p)$ in the form of 
Eq.~(\ref{texGampex}) of this paper instead of Eq.(2.12) of Ref. \cite{KO}.

Since Ohta introduced a very special dynamics, namely, an interaction
Lagrangian which generates the complete vertex function at the tree
level, he had another remarkable results. His seagull term, obtained
initially from an analysis of the $\gamma$MBB amplitude, is robust and 
appears in all electromagnetic processes which involve one or more MBB
vertex calculated with his Lagrangian. While we duplicate his result
for the $\gamma$MBB amplitude, the seagull term obtained by us will not 
appear in unmodified form in graphs involving charged meson loops. However, 
a tree graph, such as an exchange current amplitude, will have the same 
seagull term without modification. We explain the claim in the next section. 

\section{Robustness of the Seagull Term} \label{robust}
Robustness of the seagull terms is best examined in terms of subsets of
diagrams generated from a field theory satisfying the standard
requirements of invariances and symmetries.

The essence of this approach is to take a Lagrangian 
and choose  suitable but consistent subsets of diagrams for the various
physical quantities like the \se, the meson-baryon vertex function, and 
the electromagnetic amplitudes, etc and
demonstrate that the seagull terms, obtained from an analysis of the
$\gamma$MBB vertex cannot be used {\it in toto} when charged meson
loops  are  present. In principle, one example is sufficient. It is
also sufficient in practical terms if the example contains graphs which
are manifestly important.

Since we select 
diagrams we need not spell out the Lagrangian as long as there exists a 
Lagrangian which generates the diagrams we are considering. To reduce the 
number of types of diagrams we need to consider we confine ourselves to those 
which contribute to the strangeness content of the proton. Operationally,
this means that our photon couples to strangeness and not to  charge.
Therefore for the discussion of this section we have \begin{equation}
Q_P=0,\,\,\,\,Q_K=-Q_\Lambda. \label {Qlist} \end{equation}
Having chosen a subset of graphs   we do   the following: First we
define our MBB vertex function by choosing  a subset of Feynman
diagrams for it.  Then we 
couple a photon to an internal kaon or $\Lambda$ in all possible ways and 
sum them up to obtain the resulting $\gamma$MBB   vertex function. The
procedure is guaranteed to yield    vertex function which satisfies the
Ward-Takahashi identity in terms of the MBB vertex function, already
defined. The full electromagnetic 
vertex function is easily divide into two subsets - one containing  
the   pole terms, defined in section~\ref{wt}, the other constitute  
the seagull terms ${\cal M}^{sg}_{\mu}$.

We test the robustness of the seagull terms obtained from the
$\gamma$MBB vertex by constructing  
the kaon-\Lm loop which appear in the $\gamma$BB  electromagnetic 
amplitude and examining the seagull terms appearing in it to check if
they are the same as the ones from  the $\gamma$MBB vertex.  It is
straightforward to see that the seagull terms will act in a robust
manner in exchange current graphs, {\it i.e.,} tree graphs with dressed vertices.

We consider two contrasting examples. In the first example the the
KP$\Lambda$   and the  $\gamma$KP$\Lambda$ vertices arise entirely from
the proper self energy of the kaon. The baryon propagators are bare
propagators with renormalized masses. In this case we find that the
seagull terms of the $\gamma$KP$\Lambda$ vertex are robust. They appear
unmodified in the  kaon-\Lm loop term of the $\gamma$BB  electromagnetic 
amplitude . The one loop  diagrams for nucleon \se are also structurally
similar to those of Ohta. Dressed KP$\Lambda$ vertices appear at both
ends of the loop.

In the second example, the KP$\Lambda$  vertex is dressed by  a ladder
of scalar meson exchanges 
between the internal kaon and \Lm hyperon.  Here we find that while the
$\gamma$KP$\Lambda$ has the usual Ohta type seagull terms, the 
kaon-\Lm loop which appear in the $\gamma$BB  electromagnetic 
amplitude does not even have a seagull term. We also note that  the  one loop
diagrams for nucleon \se are also structurally different from those of
Ohta.  The dressed KP$\Lambda$ vertex appears at only one end of the loop.

\subsection{Kaon self-energy and the KP$\Lambda$ vertex} \label {se_ff}
We begin   with an examination of the fully dressed kaon propagator,
$\Delta(k^2)$,  which appears in the  KP$\Lambda$ vertex shown
in Fig.~\ref{figkaonp}. 
\begin{figure}[h]
\hskip 2.5in
\epsfxsize=1in
\epsfysize=0.75in
\epsffile{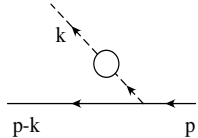}
\caption{ The KP$\Lambda$ vertex with  dressed kaon propagator, but
bare KP$\Lambda$ coupling. The circle represents kaon self-energy, not
the {\em proper} self-energy, $\Sigma(k^2)$.}
\l {figkaonp} \end{figure} 
We  write 
\beq
\Delta(k^2)=\frac{1}{k^2-m^2-\Sigma(k^2)}, 
\label{Kaonprop1}
\eeq 
where, by definition, $\Sigma(m^2)=0$, $m$ being the renormalized kaon mass. 
This allows us to write $\Sigma(k^2)=(k^2-m^2)\bar{\Sigma}(k^2)$, where 
$\bar{\Sigma}(k^2)$ is analytic in the neighborhood of $k^2=m^2$. Finally, 
we write 
\beq
\Delta(k^2)=\frac{[\tilde{f}(k^2)]^2}{k^2-m^2}, 
\label{Kaonprop2}
\eeq 
where $\tilde{f}(k^2)=[1-\bar{\Sigma}(k^2)]^{-1/2}$ is
the wave function renormalization factor in the present case.  With
$g^0_{KN\Lambda}$ as the bare coupling constant, the renormalized
coupling constant is 
\beq 
g_{KN\Lambda}=g^0_{KN\Lambda}\tilde{f}(m^2).
\label {grenorm} 
\eeq 

Amputating the baryon legs, the expression for the amplitude, shown
in Fig.~\ref{figkaonp},  is
\beq 
{\cal A}(p-k,k)= g_{KN\Lambda}f(k^2)
\frac{\tilde{f}(k^2)}{k^2-m^2},\label {kBB} 
\eeq 
where \beq f(k^2)=\frac{\tilde{f}(k^2)}{\tilde{f}(m^2)}, \label{Kff}\eeq 
is the KP$\Lambda$ form factor, normalized to $f(m^2)=1$. The last
factor in Eq.~(\ref{kBB}) is removed during amputation of the external 
propagator.

\begin{figure}[h]
\hskip 2.5in
\epsfxsize=1.2in
\epsfysize=0.9in
\epsffile{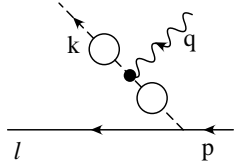}
\caption{The photo-kaon amplitude.  For brevity, graphs where the
photon couples to the $\Lambda$  have been omitted.  The baryon
propagators have been amputated.} 
\l {fig_kaonmu} 
\end{figure} 

The photo-kaon amplitude, in the present context, is exhibited in
Fig~\ref{fig_kaonmu}.  For brevity, all graphs where the photon couples
to the $\Lambda$  have been omitted.The point we wish to make can be
made without any reference to the omitted graphs. The expression for
the abbreviated amplitude  is given by  
\beq 
{\cal M}'_\mu=Q_{K}g^0_{KN\Lambda}\Delta(k^2)
\bar{\Gamma}^{(K)em}_\mu(k,k')\Delta(k'^2), \l {photoK1} \eeq 
where $\bar{\Gamma}^{(K)em}_\mu(k,k')$ is the kaon
electromagnetic vertex and  
\beq k'=k-q. \l {kprime} \eeq 
Note that the kaon electromagnetic vertex function 
$\bar{\Gamma}^{(K)em}_\mu(k,k')$ 
here is different from the $\Gamma^{(K)em}_\mu(k,k')$ as defined in 
Eq.~(\ref{eq:poledef}) in that it satisfies the Ward-Takahashi identity in
connection with the fully dressed kaon propagators
\beq
q^\mu\bar{\Gamma}^{(K)em}_\mu(k,k')=\Delta(k^2)^{-1}-
\Delta(k'^2)^{-1}.
\l {photoKWT} \eeq 
Following our practice we drop the transverse part
of $\bar{\Gamma}^{(K)em}_\mu(k,k')$ and choose for  the remainder, the 
longitudinal part, the form \cite{gross}:
\beq
\bar{\Gamma}^{(K)em}_\mu(k,k')=(k'+k)_\mu\frac{\Delta(k'^2)^{-1}-
\Delta(k^2)^{-1}}{k'^2-k^2}. \l {photoK2} \eeq
Adding and subtracting the first term 
on the right hand side of the equation below to the expression for 
${\cal M}'_\mu$, given by Eq.~(\ref{photoK1}),   and using
Eq.~(\ref{Kaonprop2})  we obtain 
\ber 
&&{\cal M}'_\mu=Q_{K}g^0_{KN\Lambda}(k'+k)_\mu[\frac{\tilde{f}(k'^2)
\tilde{f}(k^2)}{(k'^2-m^2)(k^2-m^2)}\nonumber \\
&&-\frac{\tilde{f}(k'^2)-\tilde{f}(k^2)}{k'^2-k^2}
\frac{\tilde{f}(k^2)}{k^2-m^2}-\frac{\tilde{f}(k'^2)
-\tilde{f}(k^2)}{k'^2-k^2}\frac{\tilde{f}(k'^2)}
{k'^2-m^2}]\l{photoK3} \eer
Next, we amputate the external kaon leg. This requires dividing the
expression in Eq.~(\ref{photoK3} ) by the factor
$\tilde{f}(k^2)/(k^2-m^2)$, obtaining
\ber 
&&{\cal M}''_\mu=Q_{K}g^0_{KN\Lambda}(k'+k)_\mu[\frac{\tilde{f}(k'^2) }
{k'^2-m^2 }\nonumber \\
&&-\frac{\tilde{f}(k'^2)-\tilde{f}(k^2)}{k'^2-k^2}
-\frac{\tilde{f}(k'^2)-\tilde{f}(k^2)}{k'^2-k^2}\frac{\tilde{f}
(k'^2)}{k'^2-m^2}\frac{k^2-m^2}{\tilde{f(k^2)}}]
\l{photoK4} \eer
One may verify that the second and third terms in above expression for 
${\cal M}''_\mu$ agree with the terms multiplying $Q_K$ in Eqs.~(\ref{Msg1}) 
and (\ref{mt1}) of section~\ref{wt} provided we set the KP$\Lambda$ vertex
$\bar{\Gamma}(k-q,\ell,p)=g^0_{KN\Lambda}\tilde{f}(k'^2)$.
We also note that if we had included here graphs where the photon
couples to $\Lambda$ and had also included $\Lambda$ self-energy
insertions we would have a result which matched the terms multiplying 
$Q_\Lambda$.

Finally we let $k^2\rightarrow m^2$ and 
  the last term of
Eq.~(\ref{photoK4}) drops out. Upon using Eqs~(\ref{grenorm}) and
(\ref{Kff}) the amputated amplitude becomes 
\beq 
{\cal M}_\mu=Q_{K}g_{KN\Lambda}(k'+k)_\mu[\frac{f(k'^2)}{k'^2-m^2}-
\frac{f(k^2)-f(k'^2)}{k^2-k'^2}]. \l {photoK5} 
\eeq 
The first term is the traditional generalized Born graph with dressed
KP$\Lambda$ vertex and the second term is  the , by now, familiar seagull
term of the Ohta form.

We end this subsection by noting that \bei \i[](a) because of the
simplicity of the dynamics the amplitude does not depend upon  baryon
momentum squares, \i[](b) and  that  the expression in
Eq.~(\ref{photoK3}) is fully symmetric in $k$ and $k'$. \eei
\subsection{Special dressing of the KP$\Lambda$ vertex}\label{KpLV}
We consider the set of graphs which dresses the KP$\Lambda$ vertex
with a complete set of ladders of scalar meson exchanges 
between the internal kaon and \Lm hyperon. By our definition, the
scalar meson couples to strangeness only. We call it the $\sigma$ meson
and use (gluon-like) helices for its propagators in a Feynman graph.
The complete set of ladders of $\sigma$ exchange, constituting a
K-$\Lambda$ scattering
amplitude, is shown in Fig.~\ref{fig_strip1}.
\begin{figure}[h]
\hskip 0.6in
\epsfxsize=5.2in
\epsfysize=0.88in
\epsffile{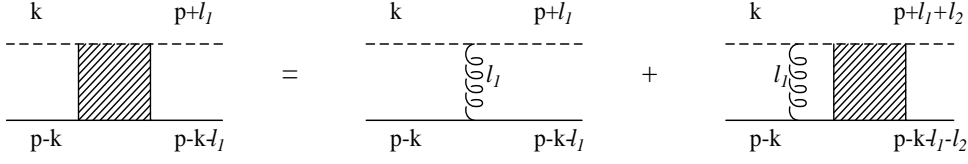}
\caption{The K-$\Lambda$ scattering amplitude. }
\l {fig_strip1} \end{figure} 
In Fig.~\ref{fig_klambda} we show the dressing of the KP$\Lambda$
vertex with a complete set of  $\sigma$-exchange ladders,
\begin{figure}[h]
\hskip 0.6in
\epsfxsize=5.2in
\epsfysize=0.8in
\epsffile{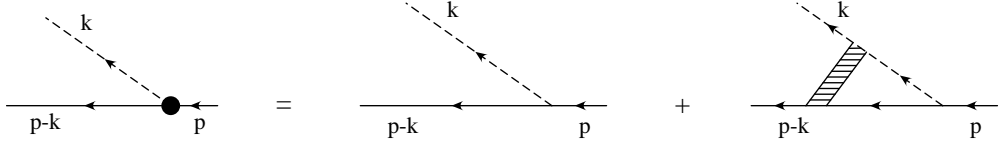}
\caption{Dressing the KP$\Lambda$ vertex with a complete set of  
$\sigma$-exchange ladders.}
\l {fig_klambda} 
\end{figure} 
\noindent
where the shaded strip represents K-$\Lambda$ scattering amplitude 
generated by a complete set of ladders of $\sigma$ exchange as shown in 
Fig.~\ref{fig_strip1}.

Notice that all self-energy insertions are absent from the chosen subset 
of diagrams.

By coupling a photon to the kaon and \Lm hyperon in all possible ways we 
generate the full photo-meson amplitude related to the $\Gamma$ vertex. 
Similar to the usage in the previous section,  $\Gamma$  represents the
vertex function corresponding to K$^{+}$ emission while $\Gamma'$ 
represents K$^{+}$ absorption (or K$^{-}$ production) vertices. 
We obtain the seagull graphs by excluding the  pole diagrams, i.e. graphs 
in which  the last interactions is the photon vertex. The resulting seagull 
graphs can be expressed by the Feynman diagrams in Fig. \ref{fig_klmu} 
\begin{figure}[h]
\hskip 0.75in
\epsfxsize=4.7in
\epsfysize=0.8in
\epsffile{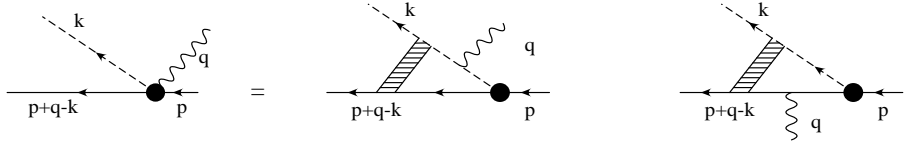}
\caption{The seagull graphs from KP$\Lambda$ vertex dressed by a complete 
set of $\sigma$-exchange ladders.}
\l {fig_klmu} 
\end{figure} 

Using the integral equations for the vertex function of Fig.
\ref{fig_klambda} and the seagull vertex 
of Fig. \ref{fig_klmu} in  an iterative procedure, it is possible to
obtain, after some patient and laborious work, the following result for
the divergence of these seagull graphs  
\begin{equation}
q^{\mu}\cdot {\cal M}^{sg}_{\mu} = 
Q_{K}\Gamma(\ell,p)+Q_{\Lambda}\Gamma(\ell-q,p). 
\label{eq:klsg}
\end{equation}
Comparing this result with Eq.~(\ref{eq:seagul}), in conjunction with
Eq.~(\ref{Qlist}),  we see that the seagull term  together with the
pole terms satisfies the \wt. Naturally, we can claim as a solution of
Eq.~(\ref{eq:klsg}) the form given by Eq.~(\ref{Msg}) with $Q_P=0$ and
$Q_\Lambda=-Q_K$.
\begin{eqnarray}
{\cal M}^{sg}_\mu &=& -Q_K\frac{(2k-q)_\mu}{k^2-(k-q)^2}
[\bar{\Gamma}(k,\ell,p)-\bar{\Gamma}(k-q,\ell,p)] \nn
&& -Q_\Lambda\frac{(2\ell-q)_\mu}{\ell^2-(\ell-q)^2}
[\bar{\Gamma}(k,\ell,p)- \bar{\Gamma}(k,\ell-q,p)] \nn
&& +Q_{\Lambda}\{\frac{(2\ell-q)_{\mu}}{\ell^2-(\ell-q)^2} 
q\!\!\!/-\gamma_{\mu}\}(g_1(k^2,(\ell-q)^2,p^2)+
g_2(k^2,(\ell-q)^2,p^2)) \nn
&& +Q_{\Lambda}\{\frac{(2\ell-q)_{\mu}}{\ell^2-(\ell-q)^2}[q\!\!\!/,p\!\!\!/]
+2i\sigma_{\mu \nu}p^{\nu}\}g_3(k^2,(\ell-q)^2,p^2). 
\label{kMsg}
\end{eqnarray} 
Thus we have produced with our choice of a subset of graphs a $\gamma$MBB 
vertex which agrees with Ohta's result.  

\subsection{Baryon self-energy and electromagnetic vertex}
\label{BEV} 
Let us consider the vertex which  measures the strangeness content of
the proton. Since the proton has strangeness zero, only loop graphs can
generate any strangeness content. The general method of constructing a
set of gauge invariant graphs is to begin with the self-energy loop
graphs and then insert one photon in all possible ways in each of these
graphs. The \wt relates the two sets of graphs.

We  follow this procedure  using 
\begin{enumerate}
\item  the Ohta prescription,
\item   the subset of graphs discussed in subsection~\ref{se_ff} and
\item    the subset of graphs discussed in 
subsection~\ref{KpLV},\end{enumerate}
and compare them.  Recall that all self-energy insertions on the
internal lines are omitted  both in the set of diagrams generated from 
the Ohta Lagrangian and from our chosen subsets of diagrams.
\subsubsection{The Ohta Approach}
In the Ohta approach the MBB vertex and the
$\gamma$MBB seagull term arise at the tree level. Hence it is quite
straightforward to write down these graphs.
\begin{figure}[h]
\hskip 2in
\epsfxsize=2in
\epsfysize=0.5in
\epsffile{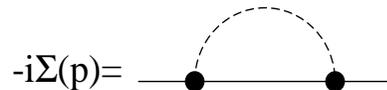}
\caption{The nucleon \se from the Ohta  prescription. }
\label{fig_otnse}
\end{figure} 
The \se graph is shown in Fig.~\ref{fig_otnse}. The proton strangeness 
vertex graphs \cite{musolf,musolf2,cohen,forkel} are shown in 
Fig.~\ref{fig_msf}. 
\begin{figure}[ht]
\hskip 0.7in
\epsfxsize=5.25in
\epsfysize=0.83in
\epsffile{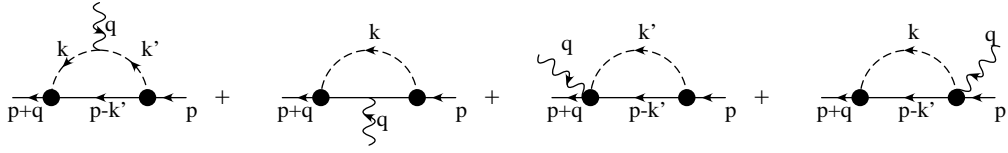}
\caption{Proton strangeness vertex according to the Ohta prescription. 
The internal lines are K  meson and $\Lambda$ or $\Sigma$. The
vertices follow from the Ohta Lagrangian at the tree level.}
\label{fig_msf}
\end{figure}
Notice the presence of the $\gamma$MBB seagull terms.

\subsubsection{Kp$\Lambda$ vertex from kaon self-energy}
\begin{figure}[h]
\hskip 1.25in
\epsfxsize=3.2in
\epsfysize=0.5in
\epsffile{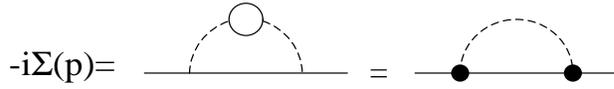}
\caption{The nucleon \se with dressed kaon propagator. The figure on
the left represents Feynman graphs, the one on the right is obtained by
using Eqs.~\protect{(\ref{grenorm}), (\ref {kBB}) and (\ref{Kff})} from
subsection~\protect{\ref{se_ff}}}
\label{fig_seff}
\end{figure}

\begin{figure}[h]
\hskip 0.7in
\epsfxsize=5.43in
\epsfysize=0.68in
\epsffile{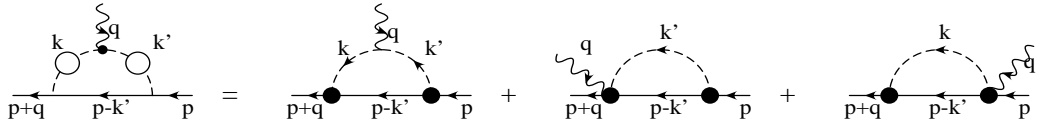}
\caption{The nucleon strangeness vertex function with dressed kaon
propagator. The graphs are identical with those in Fig.~\ref{fig_msf}
except for the second graph. As stated before this has been left out
here for the sake of brevity.}
\label{fig_svseff}
\end{figure}

The baryon self-energy graphs, in the present context, are shown in
Fig.~\ref{fig_seff}. Using Eq.~(\ref {kBB})  we find that the loop integrand 
in the left hand graph in Fig.~\ref{fig_seff} is 
$$\frac{1}{\s p -\s k -M_\Lambda}g_{KN\Lambda}f(k^2)
\frac{\tilde{f}(k^2)}{k^2-m^2}g^0_{KN\Lambda}.$$ 
Using Eqs.~(\ref{grenorm}) and  (\ref{Kff}) we rewrite the integrand as
$$\frac{1}{\s p -\s k - M_\Lambda}g_{KN\Lambda}f(k^2)
\frac{1}{k^2-m^2}f(k^2)g_{KN\Lambda},$$ which is
represented by the right hand graph. 
The proton strangeness vertex graphs for the present case are shown in
Fig~\ref{fig_svseff}.
The integrand in left hand side  graph in Fig~\ref{fig_svseff} is
$$Q_{K}\frac{1}{\s p -\s k' -M_\Lambda}g^0_{KN\Lambda}\Delta(k^2)
\bar{\Gamma}^{(K)em}_\mu(k,k')\Delta(k'^2)g^0_{KN\Lambda},$$
where $k'=k-q$. Upon using Eqs.~ (\ref{photoK2}) and (\ref{photoK3}) 
the integrand becomes
\ber 
&&Q_{K}\frac{1}{\s p -\s k' 
-M_\Lambda}g^0_{KN\Lambda}(k'+k)_\mu[\frac{\tilde{f}(k'^2)
\tilde{f}(k^2)}{(k'^2-m^2)(k^2-m^2)}\nonumber \\
&&-\frac{\tilde{f}(k'^2)-\tilde{f}(k^2)}{k'^2-k^2}
\frac{\tilde{f}(k^2)}{k^2-m^2}-\frac{\tilde{f}(k'^2)
-\tilde{f}(k^2)}{k'^2-k^2}\frac{\tilde{f}(k'^2)}{k'^2-m^2}]
g^0_{KN\Lambda}.
\eer
Finally, using Eqs.~(\ref{grenorm}) and  (\ref{Kff}) we see that the
three terms correspond to the three figures on the right hand side of
Fig~\ref{fig_svseff}. Thus in this example the seagull terms are
robust.  At the same time the self energy graphs have the same
structure as those of Ohta.
 
\subsubsection{MBB Vertex Dressed with $\sigma$ Ladders}
The \se graphs are obtained by closing the MBB vertex graphs displayed
in Fig.~\ref{fig_klambda} and the results are shown in 
\begin{figure}[h]
\hskip 1.25in
\epsfxsize=3.2in
\epsfysize=0.42in
\epsffile{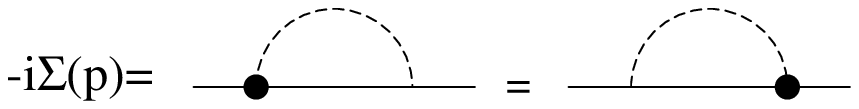}
\caption{The nucleon \se when the MBB vertex is dressed with a complete
set of $\sigma$ exchange ladders.}
\label{fig_ladse}
\end{figure}
Fig.~\ref{fig_ladse}. In sharp contrast to the diagrams arising from
the Ohta approach, as shown in Fig.~\ref{fig_otnse}, here only one MBB
vertex is dressed, the other is just the bare vertex.  The reason for
this quite obvious.  A typical ladder, shown in Fig.~\ref{fig_ladgraphs}(a), 
\begin{figure}[h]
\hskip 0.8in
\epsfxsize=4.7in
\epsfysize=0.85in
\epsffile{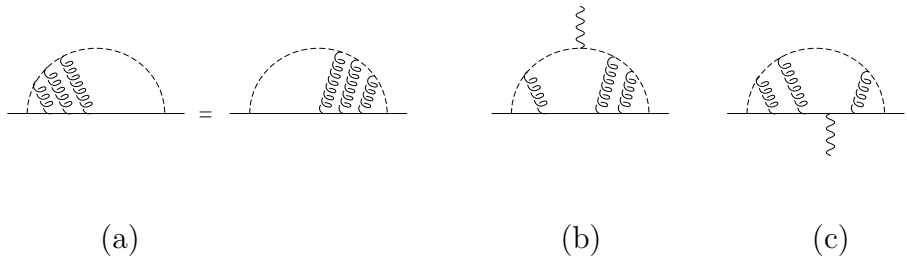}
\vskip 0.25in
\centerline{(a)\hspace{2.2in}(b)\hspace{1.1in}(c)}
\caption{A typical member of the \se graphs is shown in (a). Typical
members of the $\gamma$MBB vertex, obtained by inserting one photon
into the graph (a), are shown in (b) and (c).}
\label{fig_ladgraphs}
\end{figure} 
can be lumped as dressing of one of the two bare vertices. The ladder cannot
be split to dress both vertices.

The $\gamma$MBB vertex graphs are obtained by inserting one photon in
all possible ways in every  \se graph. Figs~\ref{fig_ladgraphs}(b) and
(c) are two examples. When we consider all such graphs, it is clear
that the ladders can be lumped into dressing the left and the right
vertices giving rise
\begin{figure}[h]
\hskip 2in
\epsfxsize=2.7in
\epsfysize=0.95in
\epsffile{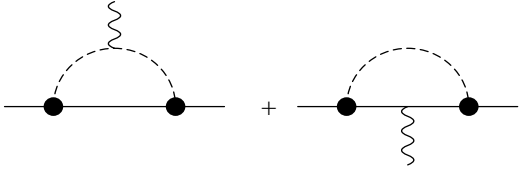}
\caption{Proton strangeness vertex diagrams following from the dressing
of the KP$\Lambda$ vertex with  a complete set of ladders of $\sigma$
exchange.  The MBB vertex is defined in Fig.~\protect\ref{fig_klambda} 
and the seagull term in Fig.~\protect\ref{fig_klmu}.}
\l {fig_nse} 
\end{figure} 
to the two dressed graphs of Fig.~\ref{fig_nse}. 
There are no seagull terms in
Fig.~\ref{fig_nse}, which is in sharp contrast to the result of the
Ohta approach described
by Fig.~\ref{fig_msf}.

In reality the KP$\Lambda$ vertex will be dressed not simply by a
complete set of ladders
 but in various other ways. Many of these will generate seagull
terms which are the same as those occurring in the corresponding 
$\gamma$MBB vertex
for the particular  form of dressing.  At the same time the dressing by
a complete set of ladders
 is manifestly an important process.  Thus we can assert that the
seagull terms which occur in the calculation of proton strangeness
content are not the same as those which occur in the $\gamma$KP\Lm vertex.  
A complete knowledge of the algebraic form
of the $\gamma$KP\Lm vertex will not enable us to predict the
seagull terms which may occur in the proton strangeness content
calculation.  In other words Ohta's prescription does not work in a
general loop graph.

\section{Results and Conclusions} \label{rescon}
It has been known for some time that baryon  electromagnetic amplitudes
have problem with gauge invariance if  meson-baryon form factors are
used. 

Recognizing these difficulties Ohta~\cite{KO} tackled the problem of 
Ward-Takahashi identity for $\gamma$MBB vertex in terms of MBB vertex
by first introducing
a dynamical basis. Given a MBB vertex, he wrote down an interaction
Lagrangian which gave the specified vertex at the tree level. The
electromagnetic interaction was introduced via minimal substitution
ensuring that the resulting $\gamma$MBB vertex would satisfy the  
Ward-Takahashi identity. The novel features of his result  was  
(a) the appearance of  a seagull term
 determined entirely by the MBB vertex and (b) that the seagull term is
{\em robust}, {\it i.e.} it appears unchanged in other electromagnetic 
amplitudes. 

It is important to note that
the transverse part of the $\gamma$MBB vertex is not fixed by this
procedure as it is not constrained by considerations of  Ward-Takahashi 
identity.

We demonstrate  that an interaction Lagrangian which generates the MBB
vertex function at tree level cannot be either hermitian or invariant
under charge conjugation. Yet we find that Ohta's discovery of the
presence of a seagull term in
the photo-meson amplitude is correct and its  algebraic relation with
the meson-baryon vertex function is also correct. Most importantly, we
find that this result is totally independent of any details of the
dynamics.  Specific knowledge of the  Lagrangian is not needed.

Unfortunately, we also find that,  in general, the resulting seagull
term is not robust.  A loop graph of an electromagnetic amplitude will
contain a seagull term if a meson baryon form factor has been used, but
it will not be the same as that appearing in a photo-meson amplitude.
This point was established with the help of a gauge-invariant subset of
graphs which describes the dressing of the meson-baryon vertex with a
complete set of ladders of $\sigma$-exchange. In this case while there
is a seagull term in the photo-meson amplitude, there is none in the 
loop graph for charge radius or strangeness content.

The seagull terms always act in a robust manner in exchange current graphs.

The problem of gauge invariance in calculations with meson-baryon
form factors is very important, particularly, in view of the proposed
experiments at the Jefferson Lab to measure the strangeness content of
proton. It is unfortunate  that Ohta's clever prescription turns out 
not to be quite correct. Obviously more work is needed to come up with
a constructive proposal to solve the problem in a gauge invariant
manner. This may require some  approximations in handling the strong
interaction dynamics.

\section*{ACKNOWLEDGMENTS}
This work was supported by DOE Grant DOE-FG02-93ER-40762.  We thank 
Tom Cohen, Michael Frank, Hilmar Forkel and Yasuo Umino for discussions.

\end{document}